\documentclass{article}
\usepackage{epsfig}
\begin{document}
\baselineskip=16pt
\begin{center}
{\Large \textbf{On the Electromagnetism of Gravitational
System and the Four Dimensional Constants ($c,\hbar, k, G$)}}\\
\vspace{2cm}
\textbf{\Large Arbab I. Arbab} \\
\vspace{1cm} {\small  Department of Physics, Teachers'
College, Riyadh 11491, P.O.Box 4341, Kingdom of Saudi Arabia \\
\emph{and} Comboni College for Computer Science, P.O. Box 114, Khartoum, Sudan}\\
E-mail: arbab@ictp.trieste.it\\
\end{center}
\vspace{1cm} {\small \textbf{Abstract}. We presented a model for
unification of electricity and gravity. We have found a consistent
description of all physical quantities pertaining to the system.
We have provided limiting values for all physical values. These
values are neither zero nor infinity. Our universe is described at
all times by the four dimensional constants $c, \hbar, k, G$ only.
The remnant of vacuum remains at all
epochs with different values. The present cosmological puzzles are justified as due to
the consequences of cosmic quantization developed in this work.}
 \vspace{.5in}
\section{Introduction}
Many attempts have failed to unify gravity with quantum mechanics.
We propose here a new approach for unification of gravity,
electricity and magnetism. It is based on the idea that the
gravito-electric effects can not be ignored at large scale. This
is achieved by defining an appropriate Planck constant that takes
care of the large scale effect. In this sense one considers the
universe to have a quantum nature present at all levels. The
quantum nature is manifested by  gravitationally bound system
only, which we call cosmic system.  General relativity predicts a
singularity at the Big Bang and within a Black Hole.  A Black Hole
is understood when quantum analysis is developed for it. We have
found that all cosmic systems require a quantum treatment as well.
So in a real quantum world such problems should not be present.
Consequently, a quantum treatment should remove the singularity
problem by allowing all physical quantities to have a limiting
values; neither zero nor infinity. The electromagnetic
contribution arising from gravitational system is very genuine and
should be taken into consideration. Such a contribution could
offset the difference between the presently observed and
anticipated energy density of the universe. This amounts to say
that the dark energy problem is no longer a problem.

The quantum nature of the whole universe is evident in the
acceleration of the cosmic fluid that permeates the space time at
different scales. And in particular, at present time there should
be a uniform acceleration of this cosmic fluid of the order of
$10^{-10}\rm m\ s^{-2}$ permeating the whole universe. Such a
value is observed in Casimir experiment and by the Pioneer
satellite.
\\
In this work we provide the limiting values of the universe at
different stages. We have written all physical quantities
representing the universe in terms of the four fundamental
constants, viz., $c, \hbar, k, G$. The universe at different
levels is governed by this set of equations. Thus the universe
appears at these stage the way it understood because it is the
only way it could. It turns out that some of the physical
quantities are relativistic, quantum, gravitational and
electromagnetic. This is evident from the way it depends on the
corresponding constants. We remark that the speed of light does
not depend of the size of the system under consideration. However,
the Planck's constant depends on the size of the system since its
unit is $ML^2T^{-1}$. So it is large for larger systems and small
for smaller systems. Hence, we expect that its value for a
macroscopic to be very large (large $M$ and $L$). The inclusion of
$G$ into the model is to represent gravity (mass), or equivalently
to represent space-time.\\
In a recent work we have found that for every bound system
(nucleus, atom, star, galaxy, the whole universe) there is a
characteristic Planck constant (Arbab, 2001a). Hence, a particular
system interacts quantum mechanically with its corresponding
Planck's constant. With this prescription all gravitational
phenomena are interpreted int terms of quantum ones. With this
remedy in  mind, the gravitational systems are successfully
described. Hence, any bound gravitational system exhibits the
quantum nature (phenomena) if it is fully understood. We have seen
that an object (mass) whether charged or not exhibits the
electromagnetic phenomena. That is because an electromagnetic
field is associated with every gravitational  bound system. Thus
an (un)charged mass in a gravitational field interacts as if it
were a charged mass placed in electromagnetic field. Its
corresponding electromagnetic fields and voltage are
$E=\left(\frac{c^7k}{\hbar G^2}\right)^{\frac{1}{2}}, \
B=\left(\frac{c^5k}{\hbar G^2}\right)^{\frac{1}{2}}$ and
$V=\left(\frac{c^4k}{ G}\right)^{\frac{1}{2}}.$ At the present
time some cosmologists believe that a new gravitational phenomena
is thought to show up in extra dimension.  An extra dimensions of
$\sim 10^{-17}\rm cm$ is thought to allow an exponential increase
in the strength of gravity to where it would match the strength of
the electro-weak and strong forces at the remarkably modest energy
of about 1 TeV. This is nearly 16 orders of magnitude below the
Planck scale, which is $10^{19}\rm GeV$, the dogmatically assumed
unification energy of all nature's forces. If the extra dimension
concept is valid, then gravity should participate equally with the
strong and electro-weak forces in the synthesis of exotic new
quanta in the supersymmetry mass range. Arguments are advanced to
support the thesis that at least one of these quanta should be
endowed with a pseudo-gravity field, whose strength is equal to
the electromagnetic field in its low energy form. We, however,
have shown in an earlier work that the gravitational constant at
Planck's time is the same as it is now (Arbab, 2001b).

Assuming all fundamental particles, with rest mass, harbor one or
more of these quanta (in virtual form) at their cores, where
vacuum tension is maximal, the result would be a short range warp
'bubble' enveloping all particles with mass. Periodic reversal of
this quanta's field would give rise to cavity oscillator behavior,
subjecting its host particle to an alternating polarity warp
metric, whose intensity would match the electromagnetic field. If
the phase of this periodicity synchronizes with the cyclical
acceleration/deceleration forces of an electron in elliptical
orbit about its nucleus, and the warp-field is always aligned
along the electron/nucleus axis, then the electron would follow a
sinusoidal, time-like geodesic through space-time, negating
synchrotron radiation. The oscillations of this natural,
micro-warp field are therefore proposed to be the essence of de
Broglie matter waves, which are the basis of stable, non-radiating
atomic orbits, and the starting point for wave mechanics.

The volumetric variations, within the warp 'bubble', are proposed
to alternate between Minkowski space and the extra dimensions,
giving rise to bi-polar relativistic sync shifts (relativity of
simultaneity), due to the resulting bi-directional linear
translations between particles. Consequently, all fundamental
particles will appear to rapidly oscillate between the past and
future at de Broglie frequencies, but average to the local
present.

For bound gravitational object the space time inside the object
and out side is different. Thus particles moving inside these
objects experience an acceleration which is different from those
moving outside. For instance, the tension of space-time inside the
nuclear region is enormously reduced, compared to the outside
tension. This may elucidate the asymptotic free nature of quarks
residing inside baryons. The surface tension of the nucleus is the
same as that of the whole universe.

In 1984 DerSarkissian suggested that a cosmic version of ordinary
quantum  mechanics  may  be responsible for the observed physical
properties of galaxies. Agob {\it et al} (1998) used a fractal
space time has shown that the Solar System is a quantized system.
They have found a cosmological Planck's constant for the galaxies
of the order of $\hbar_{\rm g}\sim 10^{67}\rm Js$. With this huge
value the expect a radio emission to dominate galaxies. A similar
form of cosmic quantum mechanics was suggested independently by
Cocke (1984). Recently  (Arbab, 2004; 2001a) we have shown that
the hierarchial problems of the matter buildup of the universe is
resolved with the idea of large scale quantization. In this work
we provide the lower and upper limits of our physical quantities.
They are neither zero nor infinity. Consequently  infinities can
not occur in our physical world, i.e., no ultraviolet no infrared
catastrophes in our theories. Thus, we notice that no fundamental
can be set to zero ($\Lambda\ne 0, \hbar\ne 0,$ etc ) as this
would violate the cosmic quantum hypothesis.

\section{The Model}
In order to unify gravity with electricity and magnetism one
requires that only fundamental constant describing these domains
should appear. These systems are described by the following
constant:
\begin{equation}\label{1}
G\ ,\ k\ , c\ , \hbar\ .
\end{equation}
With this prescription one can define the quantum effect of
gravitational and electromagnetic systems. In order for gravity to
unify with electricity they should have once had same strength.
This would mean that one had at some time the equation
\begin{equation}\label{1}
Gm^2=kq^2\ ,
\end{equation}
where $k=\frac{1}{4\pi \epsilon_0}$ is the electrical constant,
which means that gravitational and electric forces between
elementary particles with mass $m$ and charge $q$ were equal. We
argue that this force however remains unchanged (conserved). Its
value at Planck's time  and today is the same; and all other
forces are derived from it. Its value at Planck's time is
\begin{equation}\label{1}
F_P=\frac{Gm_P^2}{r_P^2}\sim 10^{43}\rm N,
\end{equation}
and its value today is
\begin{equation}\label{1}
F_0=\frac{GM_0^2}{R_0^2}\sim 10^{43}\rm N,
\end{equation}
where $R_0\sim 10^{26}\rm \ m$, $r_P\sim 10^{-35}\rm \ m $,
$m_P\sim 10^{-8}\rm \ kg$ and $M_0\sim 10^{53}\rm \ kg$. It seems
the gravitational force tiding the universe is conserved. In 1948
Casimir has found an attractive force between two plates of the
form $ F=\frac{8\pi hc}{480L^4}A$, where $A$ is the area of the
plate and $L$ is the separation between the two plates. If one
calculates this force for the present time with the cosmic
Planck's constant $(h\sim 10^{87} J.s $, see the next section) and
Planck's time, regarding the universe as a sphere, one finds the
same value. This force is attributed as due to quantum vacuum
fluctuations of the electromagnetic field. Hence, one realizes
that such a quantum nature does still exist, and has now become
sizable.

One can define a gravitational charge as
\begin{equation}\label{1}
q=\sqrt{\frac{G}{k}}\ M \ ,
\end{equation}
for a system whose gravitational mass is $M$. Moreover, we have
shown recently that the Planck constant for large scale system is
defined by (Arbab, 2001a)
\begin{equation}\label{1}
\hbar_c=\frac{GM_P^2}{c}\  ,
\end{equation}
where $M_P$ is the cosmic Planck's mass. This equation represents
a bi-pass from electric system to gravitational system. So if some
phenomena is known in one system the corresponding quantity will
be expected to take place for the other system. Space-time is
connected by strings whose tension is defined by
\begin{equation}\label{1}
T=\frac{c^4}{8\pi G}\ ,
\end{equation}
This value happens to be very huge. It implies that the space-time
is incredibly stiff ($T\sim 10^{43}\rm N$) and no stress-energy
density can make it bend no matter how big it is. The string has a
duality principle that for a physical quantity $r$ the relation
\begin{equation}\label{1}
r'=\frac{\hbar c}{r}\frac{1}{T}\ ,
\end{equation}
is also applicable. If the effective Planck's area is really
increasing ass the universe expands, that suggests the universe
will become more and more  gravitationally quantized larger
scales. The Planck's area is given by the product of the classical
gravity radius and the quantum radius as
\begin{equation}\label{1}
A=\frac{GM}{c^2}.\frac{\hbar}{M}=\left(\frac{G\hbar}{c^3}\right)\
.
\end{equation}
Now we see that this area at Planck's time ($A_{pl})$ and at the
present time $(A_0$) are  respectively
\begin{equation}\label{1}
A_{\rm P}\sim 10^{-69}\rm m^2\ ,
\end{equation}
and
\begin{equation}\label{1}
A_{0}\sim 10^{52}\rm m^2\ ,
\end{equation}
respecting the above mentioned duality.  Thus the two theories
would be applicable. We see that the same formula governs the
microscopic as well as macroscopic worlds. Therefore, the two
worlds are complementary to one another.

Now define the following physical quantities describing our system
as follows:
\subsection{Electromagnetic quantities}
These quantities provide limiting values for accessible physical
quantities. According to our hypothesis one can write this in
terms of our fundamental constant as
\begin{equation}\label{1}
\mu_B=\left(\frac{G\hbar^2}{k}\right)^{\frac{1}{2}}\ ,
\end{equation}
This is defined formally by
\begin{equation}\label{1}
\mu_B=IA\ ,
\end{equation}
where $I$ is the current flowing around the loop whose area is
$A$. Using eq.(9) the above equation yields,
\begin{equation}\label{1}
I=\left(\frac{c^6}{Gk}\right)^{\frac{1}{2}}\ .
\end{equation}
The electric field $E$ is defined as
\begin{equation}\label{1}
E=\left(\frac{c^7k}{\hbar G^2}\right)^{\frac{1}{2}}\ .
\end{equation}
The potential difference is given by
\begin{equation}\label{1}
V=\left(\frac{c^4k}{G}\right)^{\frac{1}{2}}\ .
\end{equation}
The magnetic field is defined as
\begin{equation}\label{1}
B=\left(\frac{c^5k}{\hbar G^2}\right)^{\frac{1}{2}}\ .
\end{equation}
The magnetic flux density is given by
\begin{equation}\label{1}
\Phi=\left(\frac{k\hbar}{c}\right)^{\frac{1}{2}}\ .
\end{equation}
The surface tension ( space-time stiffness constant) is defined as
\begin{equation}\label{1}
\gamma=\left(\frac{c^{11}}{\hbar G^3}\right)^{\frac{1}{2}}\ .
\end{equation}
The surface energy is given by
\begin{equation}\label{1}
U=\gamma A= \gamma\left(\frac{G\hbar}{c^3}\right)^{\frac{1}{2}}\ .
\end{equation}
The surface density is defined as
\begin{equation}\label{1}
S=\left(\frac{c^7}{\hbar
G^3}\right)^{\frac{1}{2}}=\frac{\gamma}{c^2}\ .
\end{equation}

The electric charge density is defined as
\begin{equation}\label{1}
\rho_{\rm Q}=\left(\frac{c^{10}}{\hbar^2
G^3k}\right)^{\frac{1}{2}}\ .
\end{equation}
The magnetic (electric) field contribution to mass density is
given by
\begin{equation}\label{1}
\rho_{\rm m}=\left(\frac{c^5}{\hbar G^2}\right)\ .
\end{equation}
This can be written as
\begin{equation}\label{1}
\rho_{\rm m}=\left(\frac{B^2}{k}\right)\ .
\end{equation}
The pressure is given by
\begin{equation}\label{1}
P=\left(\frac{c^7}{G^2\hbar}\right) .
\end{equation}
The amount of energy emitted per unit time per unit area (energy
flux) is given by
\begin{equation}\label{1}
\Sigma=\left(\frac{c^8}{G^2\hbar}\right) .
\end{equation}
The moment of inertia of a gravitating mass about its center is
given by
\begin{equation}\label{1}
I_c=\left(\frac{G\hbar^3}{c^5}\right)^\frac{1}{2} .
\end{equation}
The gravitational field is defined as
\begin{equation}\label{1}
\phi=\left(\frac{c^2}{G}\right)\ .
\end{equation}
The mass flow rate is defined by
\begin{equation}\label{1}
Q=\left(\frac{c^3}{G}\right)\ .
\end{equation}
The electric conductivity is defined by
\begin{equation}\label{1}
\sigma=\left(\frac{c^5}{\hbar k^2G}\right)^{\frac{1}{2}}\ .
\end{equation}
The acceleration of the quantum fluid filling the space-time is
giving by
\begin{equation}\label{1}
a=\left(\frac{c^7}{G\hbar}\right)^{\frac{1}{2}}\ .
\end{equation}
The acceleration of a charged particle (of charge $q$ and mass
$m$) in an electric field ($E$) is given by
\begin{equation}\label{1}
a=\frac{q}{m}E\ ,
\end{equation}
where $E$ is defined above. Using eq.(5) this equation yields
\begin{equation}\label{1}
a=\left(\frac{G}{k}\right)^{\frac{1}{2}}E\ ,
\end{equation}
valid for all gravitationally bound system. This acceleration
coincides with the definition
\begin{equation}\label{1}
a=\left(\frac{GM}{R^2}\right)\ ,
\end{equation}
for a gravitational system with mass $M$ and radius $R$. We remark
here the space-time (vacuum) accelerate due to its very nature.
This acceleration is required to allow the matter to be placed in
it. That is because there is a limiting mass that can be placed at
a given region. This is given by the quantity $\frac{c^2}{G}$
mentioned above. The energy embedded in this space time decays to
give the matter we observe today. However, the decay (transfer)
rate is limited to the value governed by the quantity
$\frac{c^3}{G}$. The space-time accelerate to give more space for
the created matter to be placed in. Thus space-time
(vacuum/quantum) should have a definite geometric structure. Thus
space-time represent a state of a minimum energy. Hence energy can
not be destroyed completely. The remanent of it will correspond to
space-time. The minimum energy (ground state) may not be
noticeable. But its effect can be observed by the way in which the
primeval matter is created in the universe. So this minimum energy
state would entitled a universal reference for the motion of
matter.
\\
One can define a capacitance of a gravito-electromagnetic system
as follows
\begin{equation}\label{1}
C=\left(\frac{\hbar G}{k^2c^3}\right)^{\frac{1}{2}}\ .
\end{equation}
The charge per unit length ($\lambda_q$) is given by $EC$, or
\begin{equation}\label{1}
\lambda_q=EC=\left(\frac{c^4}{Gk}\right)^{\frac{1}{2}}.
\end{equation}
One can define a diffusion coefficient (area/sec) as
\begin{equation}\label{1}
D=\left(\frac{G\hbar}{c}\right)^{\frac{1}{2}}.
\end{equation}
Using eq.(31) one finds that
\begin{equation}\label{1}
D a=c^3 \ .
\end{equation}
This means that a highly accelerating object is less diffusing,
and vice versa. One can also write the equation relating the
electrical conductivity ($\sigma$) to the coefficient of diffusion
($D$) as
\begin{equation}\label{1}
D =\left(\frac{c^2}{k\ \sigma}\right) \ .
\end{equation}
Moreover one finds that the mass and the coefficient of diffusion
are canonical conjugate to each other, i.e.,
\begin{equation}\label{1}
D M=\hbar \ .
\end{equation}
It has been emphasized by Kozlowska and Kozlowski (2003) that as
time goes one the universe becomes more and more quantum on large
scale by allowing $\hbar\rightarrow\infty$. They concluded that
the prevailing thermal process for thermal phenomena in the
universe (that taking place on large scale) is the diffusion.
\section{Planckian domain}
We calculate the above quantities at Planck's times. Now we see
that
\begin{equation}\label{1}
\mu_{\rm B_P}=\left(\frac{G\hbar^2}{k}\right)^{\frac{1}{2}}\sim
10^{-44}\rm \ J/T\ ,
\end{equation}
and
\begin{equation}\label{1}
I_{\rm P}=\left(\frac{c^6}{ Gk}\right)^{\frac{1}{2}}\sim
10^{25}\rm\ Amp\ .
\end{equation}
The Planckian electric field intensity is given by
\begin{equation}\label{1}
E_{\rm P}=\left(\frac{c^7k}{\hbar G^2}\right)^{\frac{1}{2}}\sim
10^{61}\rm\ V/m\ ,
\end{equation}
which is a typical Planckian field. The Planckian magnetic field
density is
\begin{equation}\label{1}
B_{\rm P}=\left(\frac{c^5k}{\hbar G^2}\right)^{\frac{1}{2}}\sim
10^{53}\rm\ T .
\end{equation}
The Planckian magnetic flux density is given by
\begin{equation}\label{1}
\Phi_{\rm P}=\left(\frac{k\hbar}{c}\right)^{\frac{1}{2}}\sim
10^{-17}\rm Wb .
\end{equation}
The magnetic  (electric) field contribution to mass density  is
given by
\begin{equation}\label{1}
\rho_{\rm mP}=\left(\frac{B^2}{k}\right)\sim 10^{97}\rm kg/ m^{3}\
.
\end{equation}
The surface tension of the universe at Planck's time is given by
\begin{equation}\label{1}
\gamma_{\rm P}=\left(\frac{c^{11}}{\hbar
G^3}\right)^{\frac{1}{2}}\sim 10^{78}\rm N/m .
\end{equation}
The surface density at Planck's time is defined as
\begin{equation}\label{1}
S_P=\left(\frac{c^7}{\hbar G^3}\right)^{\frac{1}{2}}\sim
10^{61}\rm kg/m^2\ .
\end{equation}
The pressure exerted at Planck's time is given by
\begin{equation}\label{1}
P_P=\left(\frac{c^7}{G^2\hbar}\right) \sim 10^{112}\rm N/m^2.
\end{equation}
The electric Planckian charge density is given by
\begin{equation}\label{1}
\rho_{\rm Q_P}=\left(\frac{c^{10}}{\hbar^2
G^3k}\right)^{\frac{1}{2}}\sim 10^{86}\rm\ C/ m^3 ,
\end{equation}
which is a enormously huge quantity. The acceleration of the
quantum fluid filling the space-time at Planck's time is given by
\begin{equation}\label{1}
a_P=\left(\frac{c^7}{G\hbar}\right)^{\frac{1}{2}}\sim 10^{51}\rm
m/ s^{2} .
\end{equation}
The amount of energy emitted per unit time per unit area (energy
flux) during Planck's time is given by
\begin{equation}\label{1}
\Sigma_P=\left(\frac{c^8}{G^2\hbar}\right)\sim 10^{120}\rm W/m^2.
\end{equation}
\section{Nuclear domain}
From an earlier work (Arbab, 2001b) we have shown that inside the
nuclear region, the Newton's constant ($G_N$) is given by
\begin{equation}\label{1}
G_N\sim10^{40}G\ .
\end{equation}
We see that planck's area inside the nuclear domain is given by
\begin{equation}\label{1}
A_N=\left(\frac{G_N\hbar}{c^3}\right)\sim 10^{-30}\rm\ m^2\ .
\end{equation}
This gives a range of about 1 fermi, that is a typical distance
for nucleons. Now we see that
\begin{equation}\label{1}
\mu_{\rm B_N}=\left(\frac{G_N\hbar^2}{k}\right)^{\frac{1}{2}}\sim
10^{-24}\rm \ J/T\ ,
\end{equation}
and
\begin{equation}\label{1}
I_{\rm N}=\left(\frac{c^6}{ G_Nk}\right)^{\frac{1}{2}}\sim
10^5\rm\ Amp\ .
\end{equation}
The nuclear electric field intensity is given by
\begin{equation}\label{1}
E_{\rm N}=\left(\frac{c^7k}{\hbar G_N^2}\right)^{\frac{1}{2}}\sim
10^{20}\rm\ V/m\ ,
\end{equation}
which is a typical nuclear field. The nuclear magnetic field
density is
\begin{equation}\label{1}
B_{\rm N}=\left(\frac{c^5k}{\hbar G_N^2}\right)^{\frac{1}{2}}\sim
10^{12}\rm\ T .
\end{equation}
The nuclear magnetic flux density is given by
\begin{equation}\label{1}
\Phi_{\rm N}=\left(\frac{k\hbar}{c}\right)^{\frac{1}{2}}\sim
10^{-17}\rm Wb .
\end{equation}
The magnetic (electric) field contribution to mass  density is
given by
\begin{equation}\label{1}
\rho_{\rm mN}=\left(\frac{B^2}{k}\right)\sim 10^{15}\rm kg/ m^{3}\
.
\end{equation}
We see that the nuclear density is independent of the number of
nucleons present.

The surface tension of a nuclear medium is given by
\begin{equation}\label{1}
\gamma_{\rm N}=\left(\frac{c^{11}}{\hbar
G_N^3}\right)^{\frac{1}{2}}\sim 10^{18}\rm N/m .
\end{equation}
The surface energy of the nucleus is given by
\begin{equation}\label{1}
U_N=\gamma\left(\frac{G\hbar}{c^3}\right)^{\frac{1}{2}}\sim
10^{-12}\rm J\sim \rm 10\ MeV .
\end{equation}
This coincides with the typical value for the binding energy per
nucleons. We would like to remark here the scale ($\lambda_{QCD})$
for quantum chromodynamics (QCD) is found o be in this range
($\lambda_{QCD}=66\pm 10 MeV$).

The gravitational field inside the nuclear region is
\begin{equation}\label{1}
\phi_N=\left(\frac{c^2}{G_N}\right)\sim 10^{-13}\rm kg/ m\ .
\end{equation}
This defines the maximal mass that can be placed inside the
nuclear gravitational field. Thus the maximal mass which can be
placed over a distance of $10^{-15}\rm m$ is $10^{-28}\rm kg$.
Therefore, the mass of the nucleus we come to know toady is the
only possible mass that the nucleus can hold. That is because the
space-time tension inside the nuclear region is exceedingly weak
(i.e. $T\sim 10^{3}\rm N$), in comparison with the tension outside
( which is $\sim 10^{43}\rm N$). The diffusion coefficient for
nuclear domain is
\begin{equation}\label{1}
D_N=\left(\frac{ G_N\hbar}{c}\right)^{\frac{1}{2}}\sim 10^{-7}\rm
m^2/s\ .
\end{equation}
Therefore, during the nuclear time the diffused area of the
nuclear constituents is $10^{-30}\rm m^2$. This is a typical area
of nuclear size. The surface density inside the Nuclear region is
defined as
\begin{equation}\label{1}
S_N=\left(\frac{c^7}{\hbar G_N^3}\right)^{\frac{1}{2}}\sim 10^2\rm
kg/m^2\ .
\end{equation}
The pressure exerted by nuclear medium(quantum) is given by
\begin{equation}\label{1}
P_N=\left(\frac{c^7}{G_N^2\hbar}\right) \sim 10^{32}\rm N/m^2.
\end{equation}
The electric nuclear charge density is given by
\begin{equation}\label{1}
\rho_{\rm Q_N}=\left(\frac{c^{10}}{\hbar^2
G_N^3k}\right)^{\frac{1}{2}}\sim 10^{26}\rm\ C/ m^{3} ,
\end{equation}
thus having the same magnitude as the nuclear mass density. This
implies that inside the nucleus both electricity and gravity
dominate. We calculate here the electric field of an electron
whose radius is $\sim 10^{-15}\rm m$. This is given by
$E=\frac{ke^2}{r}\sim 10^{20}\ \rm Vm^{-1}$ and its mass density
is $\rho=\frac{m_e}{r^3}\sim 10^{14}\rm \ kg\ m^{-3}$, and its
charge density is $\rho_e=\frac{e}{r^3}\sim\rm 10^{26}\rm C\
m^{-3}$. Comparing these values with the above data one sees that
an electron as a single system resembles a nucleus.

The nuclear tension is given by
\begin{equation}\label{1}
T_N=\frac{c^4}{8\pi G_N}\sim 10^{3}\rm N\ .
\end{equation}
This coincides with the value calculated for the quarks confined
in side hadrons. It is s thought that a quark-antiquark is made if
one tries to separate strongly interacting particles, in which
case the string tension is broken.

The acceleration of the quantum fluid filling the space-time
inside the nucleus is given by
\begin{equation}\label{1}
a_N=\left(\frac{c^7}{G_N\hbar}\right)^{\frac{1}{2}}\sim 10^{31}\rm
m/ s^{2}. \end{equation} The charge per unit length in the nuclear
region is given by
\begin{equation}\label{1}
\lambda_q=\left(\frac{c^4}{G_Nk}\right)^{\frac{1}{2}}\sim 10^{-3}
\rm C/ m.
\end{equation}
Thus, for a nuclear dimension one has a charge  of an order
$10^{-3}\times 10^{-15}\sim 10^{-18}\rm C$, which is the charge of
the nucleus.

The amount of energy emitted per unit time per unit area (energy
flux) in the nuclear region is given by
\begin{equation}\label{1}
\Sigma_N=\left(\frac{c^8}{G_N^2\hbar}\right)\sim 10^{40}\rm W/m^2.
\end{equation}
\section{Star domain}
For such a system (Globular Cluster) one has a corresponding
Planck's constant $\hbar_S\sim10^{52}\rm\ Js$. We see that
planck's area inside the star domain is given by
\begin{equation}\label{1}
A_S=\left(\frac{G\hbar_S}{c^3}\right)\sim 10^{17}\rm\ m^2\ .
\end{equation}
This gives a range of about $10^8\rm m$, that is a typical
distance for stars. Now we see that
\begin{equation}\label{1}
\mu_{\rm B_S}=\left(\frac{G\hbar_S^2}{k}\right)^{\frac{1}{2}}\sim
10^{42}\rm \ J/T\ ,
\end{equation}
and
\begin{equation}\label{1}
I_{\rm S}=\left(\frac{c^6}{ Gk}\right)^{\frac{1}{2}}\sim
10^{25}\rm\ Amp\ .
\end{equation}
The nuclear electric field intensity is given by
\begin{equation}\label{1}
E_{\rm S}=\left(\frac{c^7k}{\hbar_S G^2}\right)^{\frac{1}{2}}\sim
10^{17}\rm\ V/m\ ,
\end{equation}
which is a typical nuclear field. The nuclear magnetic field
density is
\begin{equation}\label{1}
B_{\rm S}=\left(\frac{c^5k}{\hbar_S G^2}\right)^{\frac{1}{2}}\sim
10^{9}\rm\ T .
\end{equation}
The star magnetic flux density is given by
\begin{equation}\label{1}
\Phi_{\rm S}=\left(\frac{k\hbar_S}{c}\right)^{\frac{1}{2}}\sim
10^{27}\rm Wb .
\end{equation}
The magnetic (electric) field contribution to mass  density is
given by
\begin{equation}\label{1}
\rho_{\rm mS}=\left(\frac{B^2}{k}\right)\sim 10^{8}\rm kg/ m^{3}\
.
\end{equation}
The surface tension of a star medium is given by
\begin{equation}\label{1}
\gamma_{\rm S}=\left(\frac{c^{11}}{\hbar_S
G^3}\right)^{\frac{1}{2}}\sim 10^{35}\rm N/m .
\end{equation}
The electric nuclear charge density is given by
\begin{equation}\label{1}
\rho_{\rm Q_S}=\left(\frac{c^{10}}{\hbar_S^2
G^3k}\right)^{\frac{1}{2}}\sim 0.1\rm\ C /m^{3} ,
\end{equation}
This implies that inside the stars electricity is considerable.
 The acceleration of the
quantum fluid filling the space-time inside the stars domain is
given by
\begin{equation}\label{1}
a_S=\left(\frac{c^7}{G\hbar_S}\right)^{\frac{1}{2}}\sim 10^{8}\rm
m/ s^{2} .
\end{equation}
\section{Galactic domain}
For such a system one has a Planck's constant $\hbar_G\sim
10^{68}\rm\ Js$. We see that planck's area inside the galactic
domain is given by
\begin{equation}\label{1}
A_G=\left(\frac{G\hbar_G}{c^3}\right)\sim 10^{33}\rm\ m^2\ .
\end{equation}
This gives a range of about $10^{17}\rm \ m$, that is a typical
distance for galaxies. Now we see that
\begin{equation}\label{1}
\mu_{\rm B_G}=\left(\frac{G\hbar_G^2}{k}\right)^{\frac{1}{2}}\sim
10^{58}\rm \ J/T\ ,
\end{equation}
and
\begin{equation}\label{1}
I_{\rm G}=\left(\frac{c^6}{ Gk}\right)^{\frac{1}{2}}\sim
10^{25}\rm\ Amp\ .
\end{equation}
The galactic electric field intensity is given by
\begin{equation}\label{1}
E_{\rm G}=\left(\frac{c^7k}{\hbar_G G^2}\right)^{\frac{1}{2}}\sim
10^{10}\rm\ V/m\ ,
\end{equation}
which is a typical galactic field. The galactic magnetic field
density is
\begin{equation}\label{1}
B_{\rm G}=\left(\frac{c^5k}{\hbar_G G^2}\right)^{\frac{1}{2}}\sim
10^2\rm\ T .
\end{equation}
The galactic magnetic flux density is given by
\begin{equation}\label{1}
\Phi_{\rm G}=\left(\frac{k\hbar_G}{c}\right)^{\frac{1}{2}}\sim
10^{35}\rm Wb .
\end{equation}
The magnetic  (electric) field contribution to mass  density is
given by
\begin{equation}\label{1}
\rho_{\rm mG}=\left(\frac{B^2}{k}\right)\sim 10^{-5}\rm kg/ m^{3}\
.
\end{equation}
The surface tension of a galactic medium is given by
\begin{equation}\label{1}
\gamma_{\rm G}=\left(\frac{c^{11}}{\hbar_G
G^3}\right)^{\frac{1}{2}}\sim 10^{27}\rm N/m .
\end{equation}
The electric galactic charge density is given by
\begin{equation}\label{1}
\rho_{\rm Q_G}=\left(\frac{c^{10}}{\hbar_G^2
G^3k}\right)^{\frac{1}{2}}\sim 10^{-33}\rm\ C/ m^{3} ,
\end{equation}
which is a vanishingly small quantity. The diffusion coefficient
(area/sec) for this system is
\begin{equation}\label{1}
D_G=\left(\frac{G\hbar_G}{c}\right)^{\frac{1}{2}}\sim 10^{25}\rm \
m^2/s.
\end{equation}
 This can be compared with value obtained by Agob {\it et al}., which is $1.9\times 10^{26}\rm
\ m^2/s$.

The acceleration of the quantum fluid filling the space-time
inside the galaxies is given by
\begin{equation}\label{1}
a_G=\left(\frac{c^7}{G\hbar_G}\right)^{\frac{1}{2}}\sim 10^{-1}\rm
m/ s^{2} .
\end{equation}
\section{Cosmic domain}
Here the system is described by the Planck's constant $\hbar_c\sim
10^{87}\rm \ Js$.  We see that planck's area inside the nuclear
domain is given by
\begin{equation}\label{1}
A_c=\left(\frac{G\hbar_c}{c^3}\right)\sim 10^{52}\rm\ m^2\ .
\end{equation}
This gives a range of about $10^{26}\rm m$, that is a typical
distance for our present universe. Now we see that
\begin{equation}\label{1}
\mu_{\rm B_c}=\left(\frac{G\hbar_c^2}{k}\right)^{\frac{1}{2}}\sim
10^{77}\rm \ J/T\ ,
\end{equation}
and
\begin{equation}\label{1}
I_{\rm c}=\left(\frac{c^6}{ Gk}\right)^{\frac{1}{2}}\sim
10^{25}\rm\ Amp\ .
\end{equation}
The cosmic electric field intensity is given by
\begin{equation}\label{1}
E_{\rm c}=\left(\frac{c^7k}{\hbar_c G^2}\right)^{\frac{1}{2}}\sim
1\rm\ V/m\ ,
\end{equation}
which is a typical cosmic field. The cosmic magnetic field density
is
\begin{equation}\label{1}
B_{\rm c}=\left(\frac{c^5k}{\hbar_c G^2}\right)^{\frac{1}{2}}\sim
10^{-8}\rm\ T .
\end{equation}
The cosmic magnetic flux density is given by
\begin{equation}\label{1}
\Phi_{\rm c}=\left(\frac{k\hbar_c}{c}\right)^{\frac{1}{2}}\sim
10^{44}\rm Wb .
\end{equation}
The magnetic (electric) field contribution to mass  density  is
given by
\begin{equation}\label{1}
\rho_{\rm mc}=\left(\frac{B^2}{k}\right)\sim 10^{-26}\rm kg/
m^{3}\ .
\end{equation}
The surface tension of a cosmic medium is given by
\begin{equation}\label{1}
\gamma_{\rm c}=\left(\frac{c^{11}}{\hbar_c
G^3}\right)^{\frac{1}{2}}\sim 10^{18}\rm N/m\ .
\end{equation}
The surface energy of the universe is given by
\begin{equation}\label{1}
U_c= \gamma\left(\frac{G\hbar_c}{c^3}\right)^{\frac{1}{2}}\sim
10^{70}\rm J\ .
\end{equation}
The amount of energy emitted per unit time per unit area in whole
universe is given by
\begin{equation}\label{1}
\Sigma_c=\left(\frac{c^8}{G^2\hbar_c}\right)\sim 1\rm W/m^2.
\end{equation}
The gravitational field is defined as
\begin{equation}\label{1}
\phi_c=\left(\frac{c^2}{G}\right)\sim 10^{27}\rm kg/ m\ .
\end{equation}
This defines the maximal mass that can be placed in gravitational
field. Thus the maximal mass which can be placed over a distance
of $10^{26}\rm m$ is $10^{53}\rm kg$. Therefore, the mass of the
universe we observe toady is the only possible mass that the
universe can hold.

The mass flow rate is defined by
\begin{equation}\label{1}
Q_c=\left(\frac{c^3}{G}\right)\sim 10^{35}\rm kg/ sec  .
\end{equation}
This implies the universe developed its entire mass during a time
of $10^{18}\rm sec.$ We therefore see that the universe appears
the way it is, because it is a highly constrained system.

The surface density at the present time at cosmic level is defined
as
\begin{equation}\label{1}
S_c=\left(\frac{c^7}{\hbar_c G^3}\right)^{\frac{1}{2}}\sim 1\rm
kg/m^2\ .
\end{equation}
The pressure exerted by vacuum(quantum) at the present time is
given by
\begin{equation}\label{1}
P_c=\left(\frac{c^7}{G^2\hbar_c}\right) \sim 10^{-9}\rm N/m^2.
\end{equation}
Comparing this with the Planck value one finds
\begin{equation}\label{1}
\frac{P_P}{P_c}=\left(\frac{\hbar_c}{\hbar}\right) \sim 10^{122}\
.
\end{equation}
Hence, not only the cosmological constant today is $122$ order of
magnitude, but several other cosmic quantities. One therefore
should not be  puzzled by the smallness of the cosmological
constant, but by the whole cosmic quantities as well. This is a
manifestation of a cosmic quantization of our universe at all
levels. It therefore very natural to observe these hierarchies in
our physical world. I think because of these hierarchies our
universe is unique, and without them we might not have a universe
lasting for 10 - 15 billion of years!

The electric cosmic charge density is given by
\begin{equation}\label{1}
\rho_{\rm Q_c}=\left(\frac{c^{10}}{\hbar_c^2
G^3k}\right)^{\frac{1}{2}}\sim 10^{-36}\rm\ C /m^{3} ,
\end{equation}
Again, this implies that the present universe can't be dominated
by electricity today.
 The acceleration of the quantum fluid filling the space-time at present's time is given by
\begin{equation}\label{1}
a_c=\left(\frac{c^7}{G\hbar_c}\right)^{\frac{1}{2}}\sim
10^{-10}\rm\ m/ s^{2} .
\end{equation}
We therefore expect all objects to have experienced a uniform
acceleration due to expansion of the cosmic fluid filling the
whole universe. Thus every object will experience this
acceleration as far as it floats on space-time. However, such an
acceleration is observed in Casimir experiments and recently
observed by Pioneer satellite. The diffusion coefficient for the
cosmic domain is given by
\begin{equation}\label{1}
D_c=\left(\frac{ G\hbar_c}{c}\right)^{\frac{1}{2}}\sim 10^{34}\
\rm\ m^2/s\ .
\end{equation}
Therefore, during the cosmic time the diffused area of the cosmos
constituents is $10^{52}\rm m^2$. This is a typical area of cosmic
size. It has been emphasized by Kozlowska and Kozlowski (2003)
that as time goes one the universe becomes more and more quantum
on large scale by allowing $\hbar\rightarrow\infty$. They
concluded that the prevailing thermal process for thermal
phenomena in the universe (that taking place on large scale) is
the diffusion.
\section{The hierarchial Universe}
We see from our analysis that the hierarchial structure of our
universe is due to that fact that the present cosmic quantities
are related by the Planckian ones by a factor that depends on
planck constants of the two systems. This factor takes into
account the smallness and the vastness of the atomic and cosmic
realms when compared to each other. And since $G_P=G_0$, one
defines this factor as:
\begin{equation}
D=\sqrt{\left(\frac{\hbar_c}{\hbar}\right)}\ \ \sim 10^{61}\ \ \ ,
\end{equation}
so that the mass, density, acceleration and pressure of the
universe are
\begin{equation}\label{1}
M_0=\left(D\right)M_P,\qquad\
\rho_0=\left(\frac{1}{D^2}\right)\rho_P,\qquad \
a_0=\left(\frac{1}{D}\right)a_P,\qquad
P_0=\left(\frac{1}{D^2}\right)P_P,
 \end{equation}
 and; the cosmological constant, radius, age and electric field of
 the universe are
\begin{equation}
\Lambda_c=\left(\frac{1}{D^2}\right)\Lambda_P, \qquad
R=\left(D\right) R_P,\qquad t_0=\left(D\right) t_P, \qquad
E_0=\left(\frac{1}{D}\right)E_P\ \ .
\end{equation}
Hence, the present cosmic quantities are big(or small) when
measured in Planckian units.
\section{References}
Agob, M. {\it et al}., \emph{Aust. J.  Phy. V.51,9, 1998}.\\
Arbab, A.I., \emph{Gen. Rel. Gravit. V.39,  2004(in press)}.\\
Arbab, A.I., \emph{Spacetime \& Substance, V.2, 55, 2001a}.\\
Arbab, A.I., \emph{Spacetime \& Substance, V.2, 51, 2001b}.\\
DerSarkissian, M., \emph{Lett. Nouvo Cimento, V.40, 390, 1984}.\\
Cocke, W.J., \emph{Astrophys. J. Lett. V.23, 239, 1983}.\\
Kozlowska, J.M., and Kozlowski, M.,
http://lanl.arxiv.org/abs/astro-ph/0307168; \emph{Foundations of
Physics Letters V.9, 235, 1996.}\\
\end{document}